\documentclass[12pt,preprint]{aastex}

\shorttitle{Extended, luminous globular clusters in M31}
\shortauthors{Mackey et al.}

\begin{document}

\title{ACS photometry of extended, luminous globular clusters in the outskirts of M31\altaffilmark{1}}


\author{A.D. Mackey\altaffilmark{2}, A. Huxor\altaffilmark{3}, A.M.N. Ferguson\altaffilmark{2}, N.R. Tanvir\altaffilmark{4}, M. Irwin\altaffilmark{5}, R. Ibata\altaffilmark{6}, T. Bridges\altaffilmark{7}, R.A. Johnson\altaffilmark{8}, G. Lewis\altaffilmark{9}}


\altaffiltext{1}{Based on observations made with the NASA/ESA Hubble Space 
Telescope, obtained at the Space Telescope Science Institute, which is operated 
by the Association of Universities for Research in Astronomy, Inc., under NASA 
contract NAS 5-26555. These observations are associated with program 10394.}
\altaffiltext{2}{Institute for Astronomy, University of Edinburgh, Royal Observatory, 
Blackford Hill, Edinburgh, EH9 3HJ, UK}
\altaffiltext{3}{Department of Physics, University of Bristol, Tyndall Avenue, 
Bristol, BS8 1TL, UK}
\altaffiltext{4}{Department of Physics \& Astronomy, University of Leicester, 
Leicester, LE1 7RH, UK}
\altaffiltext{5}{Institute of Astronomy, University of Cambridge, Madingley Road, 
Cambridge, CB3 0HA, UK}
\altaffiltext{6}{Observatoire de Strasbourg, 11 rue de l'Universit\'{e}, F-67000 Strasbourg, France}
\altaffiltext{7}{Department of Physics, Queen's University, Kingston, ON, K7M 3N6, Canada}
\altaffiltext{8}{Oxford Astrophysics, Denys Wilkinson Building, Keble Road, Oxford, 
OX1 3RH, UK}
\altaffiltext{9}{Institute of Astronomy, School of Physics, A29, University of Sydney, NSW 2006, Australia}


\begin{abstract}
A new population of extended, luminous globular clusters has recently been 
discovered in the outskirts of M31. These objects have luminosities typical
of classical globular clusters, but much larger half-light radii. We report 
the first results from deep ACS imaging of four such clusters, one of which 
is a newly-discovered example lying at a projected distance of $\sim 60$ kpc 
from M31. Our F606W, F814W colour-magnitude diagrams extend $\sim 3$ magnitudes 
below the horizontal branch level, and clearly demonstrate, for the first time, 
that all four clusters are composed of 
$\ga 10$ Gyr old, metal-poor stellar populations. No evidence for multiple 
populations is observed. From a comparison with Galactic globular cluster 
fiducials we estimate metallicities in the range $-2.2 \la [$Fe$/$H$] \la -1.8$. 
The observed horizontal branch morphologies show a clear second parameter effect 
between the clusters. Preliminary radial luminosity profiles suggest integrated 
magnitudes in the range $-6.6 \la M_V \la -7.7$, near the median value of the globular
cluster luminosity function. Our results confirm that these four 
objects are {\it bona fide} old, metal-poor globular clusters, albeit with 
combined structures and luminosities unlike those observed for any other 
globular clusters in the Local Group or beyond.
\end{abstract}




\section{Introduction}
As the nearest large spiral galaxy to our own, M31 is a prime target for
stellar populations studies aimed at understanding the formation and evolution
of such galaxies. An important subset of this work concerns the globular
cluster system of M31, since the objects belonging to this system allow us
to trace the star formation history, chemical evolution, and mass distribution
of M31. Recently, as part of a major survey of the M31 halo with the
Isaac Newton Telescope (INT), a search for previously unknown globular clusters, 
at large projected radii from the center of M31, has been conducted. Details of 
the INT survey may be found in \citet{ferguson:02} (see also the references
therein); spatial coverage now extends to more than $40$ deg$^2$. The
globular cluster search has resulted in the discovery of a significant number
of new clusters, with projected radii ($R_p$) extending to $\sim 100$ kpc 
\citep[see e.g.,][]{huxor:04,huxor:06}. Hubble Space Telescope (HST) 
imaging of many of these new clusters has recently been carried out -- first results 
from this work are described elsewhere \citep[][hereafter Paper II]{mackey:06}.

One surprising result to arise from the globular cluster survey
was the discovery of a previously unknown population of extended, luminous
objects in the outskirts of M31 \citep{huxor:05}. Specifically, three
such clusters were discovered, with $15 \la R_p \la 35$ kpc.
These clusters are notable because they have large half-light radii ($r_h$), but 
unlike known globular clusters with similar structures, also have luminosities
near the peak of the globular cluster luminosity function (GCLF). As such, they are
dissimilar to any other known clusters in the Milky Way, M31, or anywhere else, and
begin to fill in the gap in parameter space between classical globular clusters
and dwarf spheroidal galaxies \citep{huxor:05}.

Subsequent to the discovery of these three clusters, a further survey of the
southern quadrant of the outer M31 halo, and down the minor axis of M31 to M33,
has been conducted with MegaCam on the Canada-France-Hawaii
Telescope \citep[see e.g.,][]{martin:06}. An additional extended cluster 
was located with these observations, at a projected radius of $\sim 60$ kpc.
Following the nomenclature of \citet{huxor:05}, we label this object EC4.
Its coordinates are listed in Table \ref{t:results}. This is one of the most 
remote M31 globular clusters currently known, after two of the newly-discovered 
classical globular clusters at $R_p \sim 78$ and $100$ kpc 
(see Paper II), and the new cluster of \citet{martin:06}, at $R_p \sim 116$ kpc.

In this Letter, we report on the first results from imaging of the four unique
luminous, extended M31 clusters with the Advanced Camera for Surveys (ACS) on-board 
HST. 

\section{Observations and data reduction}
\begin{figure}
\epsscale{0.8}
\plotone{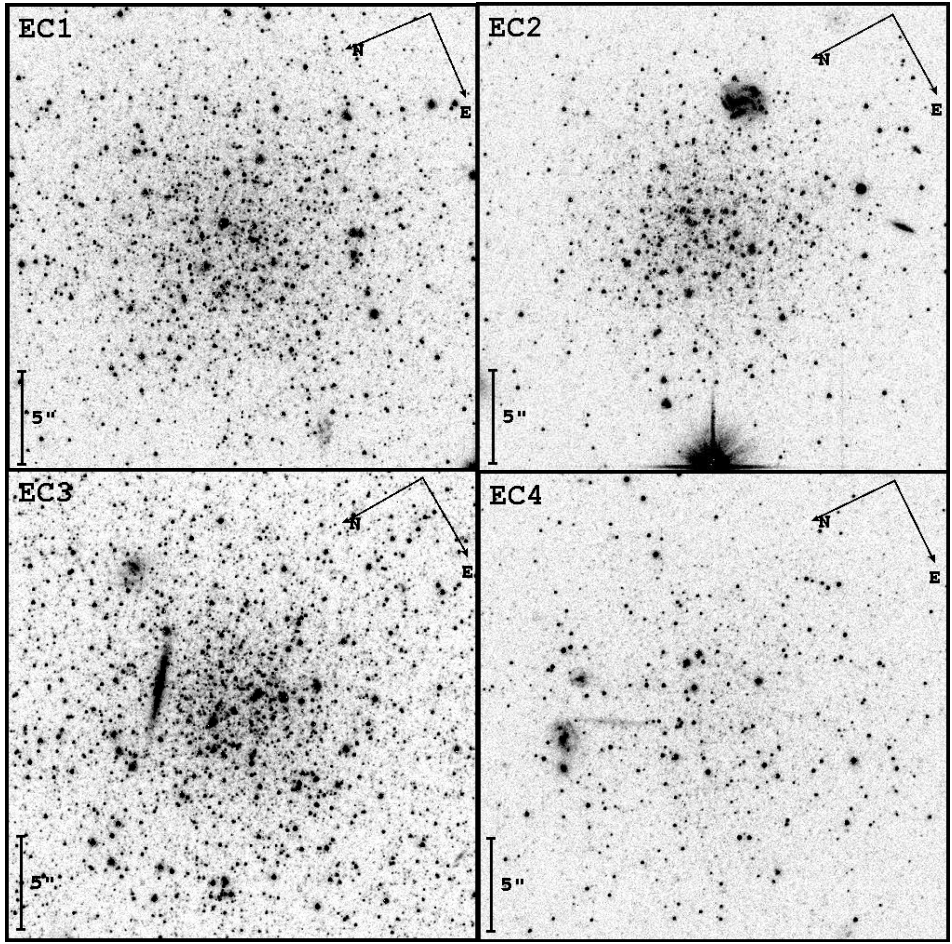}
\caption{Drizzled ACS/WFC F606W images of the four extended luminous M31 globular 
clusters. Each thumbnail has dimensions of 
$25\arcsec \times 25\arcsec$.\label{f:images}}
\end{figure}

Our observations were obtained with the ACS Wide Field Channel (WFC) under 
HST program GO 10394 (P.I. Tanvir), between 2005 July 10 -- 2005 July 23. 
Each cluster was imaged three times in the F606W filter and four times in the 
F814W filter, with small dithers between subsequent images. Typical total 
integration times were 1800s in F606W and 3000s in F814W. Each cluster was 
placed at the centre of either chip 1 or chip 2 on the WFC, to ensure that the 
inter-chip gap did not impact the observations. In Fig. \ref{f:images} we
display drizzled F606W images of the four clusters, confirming the nature of
these objects. Each is clearly very extended and of low surface brightness, 
quite different in structure to the classical globular clusters which have 
previously been observed in M31 \citep[see e.g., the images in][]{rich:05}.

We used the {\sc dolphot} photometry software \citep{dolphin:00}, 
specifically the ACS module, to photometer our images. {\sc dolphot} performs
point-spread function (PSF) fitting using PSFs especially tailored
to the ACS camera. For a given cluster, photometry was done simultaneously
on all the flat-fielded images from the STScI archive (both filters), 
relative to some deep reference image -- we used the drizzled combination 
of our F814W images. {\sc dolphot} accounts for the hot-pixel 
and cosmic-ray masking information provided with each flat-fielded image, 
fits the sky locally around each detected source, and provides magnitudes 
corrected for charge-transfer efficiency (CTE) effects on the calibrated 
VEGAMAG scale of \citet{sirianni:05}. A variety of quality information is 
listed with each detected object, including the object type (stellar, 
extended, etc), $\chi^2$ of the PSF fit, sharpness and roundness of the 
object, and a ``crowding'' parameter which describes how much brighter an 
object would have been had neighbouring objects not been fit 
simultaneously. We used some of this information to select only stars 
with high quality measurements. Specifically, we selected objects of 
stellar type, with valid photometry on all seven input images, global 
sharpness parameter between $-0.3$ and $0.3$ in each filter, and crowding 
parameter less than $0.25$ in each filter. 

We isolated each cluster from the surrounding field population by imposing a
radial cut. Two (EC1, EC3) are located, in projection, 
fairly close to the main body of M31, and hence have heavy background fields.
For these objects, we set the selection radii to be small 
($15\arcsec$ and $8\arcsec$, respectively). We found these sufficient 
to obtain clean cluster colour-magnitude diagrams (CMDs) with plenty of stars 
still available to define the primary sequences. For the two clusters at large 
projected radii (EC2, EC4), we used limiting radii closer to the 
expected cluster tidal radii, at $\sim 25-30\arcsec$. 

\section{Analysis}
\begin{figure}
\plotone{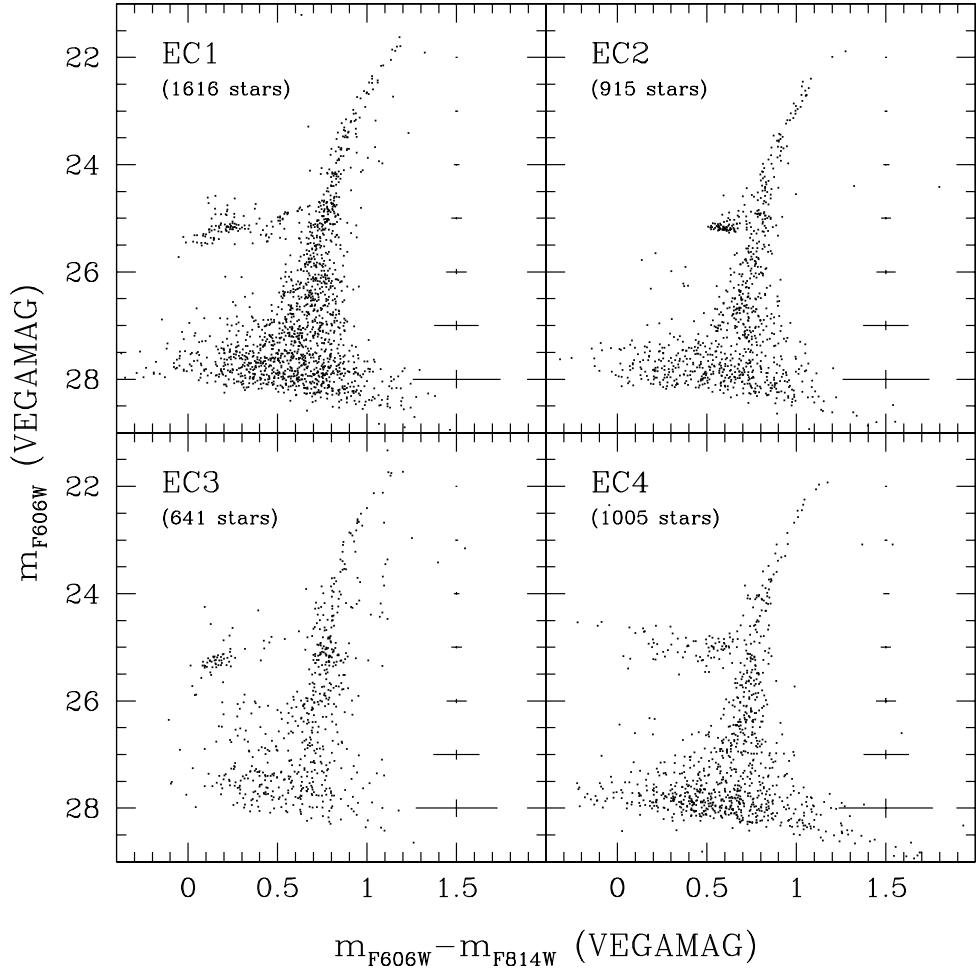}
\caption{CMDs for the four extended, luminous M31 globular clusters. Photometry has been selected from the full ACS/WFC fields by imposing radial limits of, respectively, $15\arcsec$, $25\arcsec$, $8\arcsec$, and $30\arcsec$ from the cluster centers. Typical photometric errors from {\sc dolphot} are indicated.\label{f:cmds}}
\end{figure}

Fig. \ref{f:cmds} shows CMDs for the four clusters. Our photometry 
reaches $\sim 3$ mag below the level of the horizontal branch (HB), to 
a limiting magnitude of $m_{{\rm F606W}} \sim 28$. Each cluster exhibits 
a narrow, steep red-giant branch (RGB), and clearly delineated horizontal 
branch. Three of the clusters possess HBs extending to the blue, including 
broadened regions with colours in the range 
$0.2 \la m_{{\rm F606W}} - m_{{\rm F814W}} \la 0.5$ which are
suggestive of RR Lyrae stars imaged at random points during their cycle. 
Taken together, these features are indicative of old ($\ga 10$ Gyr), metal-poor 
stellar populations. 

We observe no sign of young or intermediate-age 
populations in these clusters, such as a blue plume or luminous AGB stars above 
the RGB tip. Comparing CMDs for the inner and outer parts of each cluster 
does not reveal any radially dependent differences such as those commonly 
seen in dSph galaxies \citep[e.g.,][]{harbeck:01}. 
Furthermore, the widths of the four RGBs above the HB level are comparable to 
each other and to those for our classical globular clusters (see Paper II), and are 
consistent with the 
photometric errors. Hence, we see no evidence for multiple stellar populations 
within our four clusters. This argues against merger formation scenarios 
except those in which the parent clusters were of similar age and metallicity,
as well as scenarios involving stripped dSphs.

In order to obtain photometric metallicity estimates from our CMDs,
we used the cluster fiducial sequences in the ACS/WFC F606W and F814W bands 
presented by \citet{brown:05}. Because our four targets are distributed across
a large area of sky and a wide variety of projected radii from M31, it 
is prudent not to assume a single distance modulus or reddening value for all.
Rather, we adopted a procedure to simultaneously constrain the 
metallicity, distance modulus, and reddening for each cluster.
We chose to register the fiducials to our cluster CMDs using the F606W magnitude 
of the HB and the colour of the red-giant branch (RGB) at 
the HB level. If the metallicity of the reference fiducial matches that for 
a given cluster, then when registered using these two points, 
the fiducial will closely trace the upper RGB of the cluster. If the
reference fiducial is of incorrect metallicity, the fiducial and CMD will
deviate on the upper RGB. Bounding the cluster RGB with more metal-rich
and metal-poor fiducials allows the true cluster metallicity to be estimated.

The absorption in each ACS bandpass is a non-linear function of both 
effective temperature and $E(B-V)$. \citet{brown:05} therefore provide
two coefficients ($\alpha,\,\beta$) per bandpass at each point on their 
fiducial sequences, allowing the absorption to be calculated given $E(B-V)$. 
For example, in F606W the absorption $A_{{\rm F606W}}$ is:
\begin{equation}
A_{{\rm F606W}} = \alpha_{{\rm F606W}} E(B-V) + \beta_{{\rm F606W}} [E(B-V)]^2,
\label{eq1}
\end{equation}
and similarly in F814W. Following \citet{mcconnachie:05}, M31 has distance modulus 
$(m-M)_0 = 24.47 \pm 0.07$; however, due to the line-of-sight depth of the system, 
the cluster distance moduli must be scattered about this value. We decided to test 
over a range of $\pm 0.5$ mag, representing a system depth of $\sim 360$ kpc. For 
a given $(m-M)_0$, we used Eq. \ref{eq1} to solve for the $E(B-V)$ which aligned 
the HB levels of the observed cluster and the fiducial under consideration. With 
this constrained, we calculated the difference between the colour of the cluster 
RGB at HB level, and that of the transformed fiducial; the best solution for 
$(m-M)_0$ and $E(B-V)$ was the combination which minimised this offset. 
We derived uncertainties in these quantities by randomly selecting new HB levels for 
the fiducial and cluster and a new RGB colour for the cluster, from Gaussian 
distributions about our measured values (typical distribution widths were 
$\pm 0.1$ mag in the HB levels, and $\pm 0.01$ mag in the RGB colour), and solving 
for new $(m-M)_0$ and $E(B-V)$. Iterating ten thousand times built up distributions
from which we derived standard random errors. These are generally $0.14$ mag in 
$(m-M)_0$ and $0.01$ mag in $E(B-V)$. We note that there are possibly 
additional systematic errors present for our measured $E(B-V)$ values, due to 
differences in age between the M31 clusters and the fiducial clusters, and the
fact that the colour of the RGB at the HB level is mildly age-sensitive. At most, 
these errors would amount to $\sim 0.02$ mag more than quoted in Table \ref{t:results},
corresponding to the interval $10-15$ Gyr.

\begin{figure}
\plotone{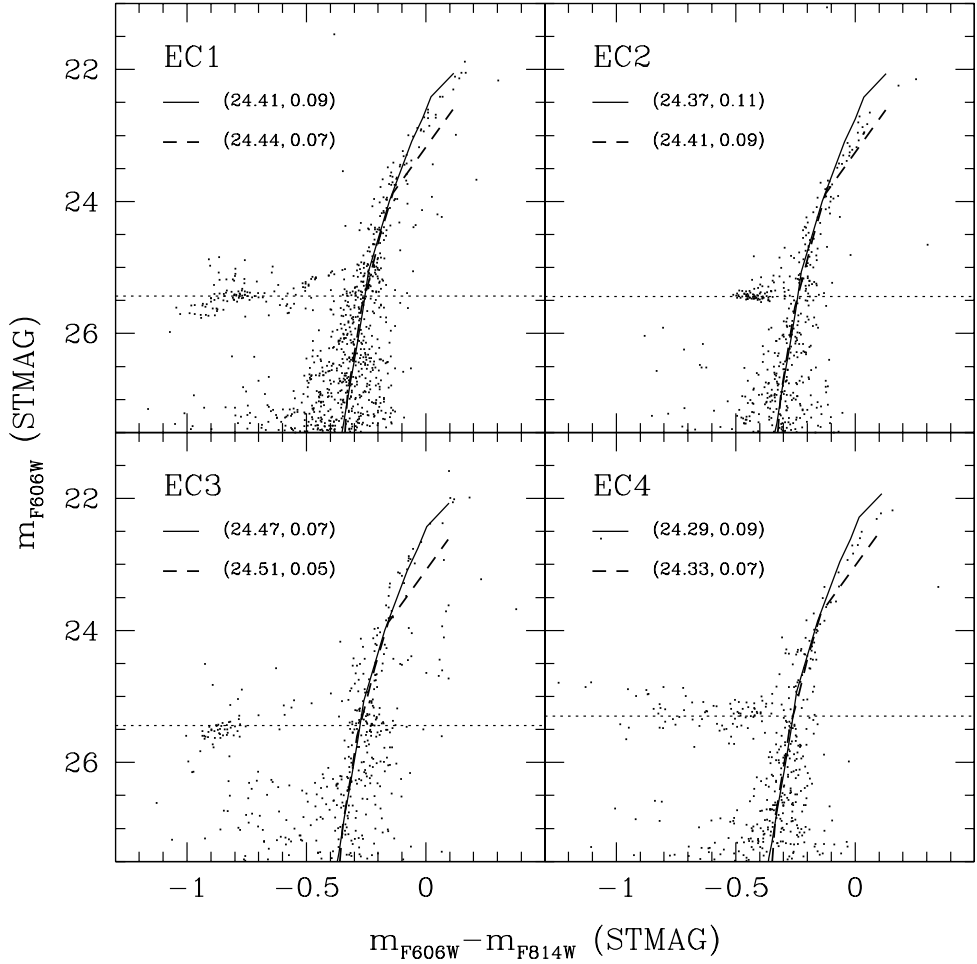}
\caption{Results of fitting the Galactic globular cluster fiducials from \citet{brown:05} to the observed CMDs. All four clusters are metal-poor, and hence have RGBs bracketed by that for NGC 6341 (M92) at $[$Fe$/$H$]=-2.14$ (solid lines), and that for NGC 6752 at $[$Fe$/$H$]=-1.54$ (dashed lines). EC3 is the most metal poor cluster, with the M92 fiducial providing a close fit. The horizontal dotted lines represent the adopted HB levels. The numbers in the upper left indicate the best fitting $(m-M)_0$ and $E(B-V)$ for the marked fiducials. Photometry has been transformed to STMAG to match the \citet{brown:05} fiducials.\label{f:fids}}
\end{figure}

With the fiducials aligned, it was straightforward to infer the most likely value 
of $[$Fe$/$H$]$ (and hence the best values for $(m-M)_0$ and $E(B-V)$) for a given 
cluster. The fiducial alignments are presented in Fig. \ref{f:fids}, and the
numerical results in Table \ref{t:results}. Typical errors in $[$Fe$/$H$]$ are
approximately $\pm 0.15$ dex. From Fig. \ref{f:fids}, it is clear that all four
clusters are metal-poor objects. Three have an RGB locus intermediate between
those for NGC 6341 (M92) and NGC 6752, while the fourth (EC3) matches closely
that for NGC 6341. \citet{brown:05} adopt $[$Fe$/$H$] = -2.14$ and $-1.54$ for 
these two clusters, respectively; the listed metallicities in Table \ref{t:results} 
reflect these values.

For comparison, we consider values of $E(B-V)$ from the extinction maps of 
\citet{schlegel:98} (see Table \ref{t:results}). The agreement 
is close, indicating the validity of our method. We note a tendency 
to over-estimate $E(B-V)$ by $0.01-0.02$ mag; \citet{barmby:00} found a similar 
effect in their large study of M31 globular clusters. Adopting $E(B-V)$ from 
\citet{schlegel:98} tends to increase $(m-M)_0$ by a few hundredths of a magnitude
and does not noticeably affect the derived $[$Fe$/$H$]$; however the alignment
for EC4 is noticeably inferior. Our derived distance moduli for the two clusters 
with smallest $R_p$ are $\sim 24.45$, matching the M31 value from 
\citet{mcconnachie:05}. EC4 may lie somewhat closer to us than the bulk of M31; 
however this is unsurprising given its large $R_p$.

From our CMDs and estimated $[$Fe$/$H$]$, it is
clear that there is a strong second parameter effect among this set of clusters.
Deriving accurate HB-morphology parameters requires variability information,
completeness tests, and a treatment of photometric blends at the red end
of the HB \citep[see e.g.,][]{rich:05}, and is therefore beyond the scope of the
present letter. We defer such calculations to a future paper, currently in 
preparation. Even so, examining the three clusters with very similar 
$[$Fe$/$H$] \sim -1.84$, the second parameter effect is easily discernible by eye. 
EC2 has an exclusively red HB, while EC4 has an intermediate HB and EC1 has a mostly 
blue HB. The CMD for EC2 is strikingly similar to that for the SMC globular NGC 121, 
which has comparable $[$Fe$/$H$]$. This cluster is $\sim 2-3$ Gyr younger than the 
oldest Galactic and LMC globular clusters \citep[see e.g.,][]{shara:98}, suggesting 
that a similar age spread may be present among our M31 clusters. 

Finally, we obtained estimates of structural parameters and total
luminosities by means of integrated (aperture) photometry.
This technique ensures unresolved cluster light is counted, and obviates any
present need for lengthy completeness tests. Using the CMDs we
estimated the largest apertures ($r_{{\rm max}}$) within which field star 
contamination does not add significantly to the integrated cluster luminosity. 
For a given cluster we used this 
aperture to obtain the total luminosity in F606W and F814W (background 
galaxies and bright field stars were masked), then used these to
transform the F606W value into Johnson $V$ using the prescription in 
\citet{sirianni:05}. We then calculated $M_V$ using our measured $E(B-V)$ and 
$(m-M)_0$. These luminosities are lower limits, since $r_{{\rm max}}$ is almost 
certainly smaller than the tidal radius in all clusters. For EC2 and EC4 the amount 
of missing light is only $\sim 0.1$ mag, since $r_{{\rm max}} \sim 75$ pc. 
However, for EC1 and EC3, which have small $r_{{\rm max}}$, the shortfall 
is significant. Based on profiles for EC2 and EC4, the true luminosity 
of EC1 may be $\sim 0.3$ mag brighter than listed, and that for EC3 $\sim 0.7$ 
mag brighter.

With $M_V$ calculated, we sampled the luminosity profile of
each cluster with a variety of apertures to determine the half-light radius $r_h$.
Subtracting the luminosities within consecutive apertures allowed preliminary
surface brightness profiles to be constructed. From each, the core radius 
$r_c$ (at which the surface brightness is half its central value) was estimated. 
Table \ref{t:results} lists $r_c$ and $r_h$, converted to parsecs using our
measured $(m-M)_0$. Typical uncertainties are $\pm 15\%$ in each.

Taken together, the above results confirm the original conclusion of \citet{huxor:05} 
that these four objects strongly resemble classical globular clusters in terms
of their old, metal-poor stellar populations, but are quite different structurally.
Specifically, their luminosities sit around the peak of the GCLF, and as such 
they are much brighter than similarly extended objects found in the Milky Way 
\citep[including the new cluster of][]{belokurov:06}, the LMC, or externally 
\citep[such as the ``faint fuzzies'' of][]{brodie:02}. They thus encroach on the 
gap in parameter space between classical globular clusters (which are dark matter
free) and dwarf spheroidal galaxies (which are dark matter dominated). Further 
study of this unique population will shed light on the transition region between 
globular clusters and dwarf spheroidals, and provide important clues to the formation 
and evolution of these clusters, which may be strongly linked to that of M31.

\acknowledgments
ADM and AMNF are supported by a Marie Curie Excellence Grant from the European 
Commission under contract MCEXT-CT-2005-025869. NRT acknowledges financial
support via a PPARC Senior Research Fellowship.



{\it Facilities:} \facility{HST (ACS)}.


\clearpage

\begin{deluxetable}{lccccccccccc}
\tabletypesize{\footnotesize}
\rotate
\tablecaption{Observed properties of four luminous, extended globular clusters in the outskirts of M31\label{t:results}}
\tablewidth{0pt}
\tablehead{
\colhead{Identifier\tablenotemark{a}} & \colhead{RA} & \colhead{Dec} & \colhead{$R_p$} & \colhead{$(m-M)_0$} & \colhead{$E(B-V)$} & \colhead{$E(B-V)$} & \colhead{$[$Fe$/$H$]$} & \colhead{$r_c$} & \colhead{$r_h$} & \colhead{$M_V$\tablenotemark{b}} & \colhead{$r_{{\rm max}}$} \\
 & \colhead{(J2000.0)} & \colhead{(J2000.0)} & \colhead{(kpc)} & & \colhead{(meas.)} & \colhead{(lit.)} & & \colhead{(pc)} & \colhead{(pc)} & & \colhead{($\arcsec$)}}
\startdata
EC1 & $00^{{\rm h}} 38^{{\rm m}} 19\fs5$ & $+41\degr 47\arcmin 15\farcs0$ & $13.3$ & $24.42 \pm 0.15$ & $0.08 \pm 0.01$ & $0.07$ & $-1.84$ & $20.4$ & $35.4$ & $-7.4$ & $16$ \\
EC2 & $00^{{\rm h}} 42^{{\rm m}} 55\fs1$ & $+43\degr 57\arcmin 28\farcs5$ & $36.8$ & $24.39 \pm 0.14$ & $0.10 \pm 0.01$ & $0.09$ & $-1.84$ & $13.2$ & $29.5$ & $-7.0$ & $20$ \\
EC3 & $00^{{\rm h}} 38^{{\rm m}} 04\fs5$ & $+40\degr 44\arcmin 37\farcs5$ & $14.0$ & $24.47 \pm 0.14$ & $0.07 \pm 0.01$ & $0.07$ & $-2.14$ & $15.1$ & $32.3$ & $-7.0$ & $9$ \\
EC4 & $00^{{\rm h}} 58^{{\rm m}} 15\fs5$ & $+38\degr 03\arcmin 01\farcs1$ & $60.0$ & $24.31 \pm 0.14$ & $0.08 \pm 0.01$ & $0.05$ & $-1.84$ & $23.8$ & $33.7$ & $-6.6$ & $20$ \\
\enddata
\tablenotetext{a}{Nomenclature follows that of \citet{huxor:05}, with the addition of a new cluster, EC4.}
\tablenotetext{b}{Integrated luminosities are lower limits, as discussed in the text. In particular, EC1 and EC3 may be $\sim 0.3$ and $\sim 0.7$ mag brighter than listed, respectively.}
\end{deluxetable}

\clearpage

\clearpage

\clearpage

\end{document}